\documentclass[12pt,preprint]{aastex}

\begin{document}

\title{Synchronized formation of starburst
and poststarburst galaxies in  merging  clusters of galaxies}

\author{Kenji Bekki} 
\affil{
ICRAR,
M468,
The University of Western Australia
35 Stirling Highway, Crawley
Western Australia, 6009
}

\and

\author{Matt S. Owers and
Warrick J. Couch}
\affil{
Centre for Astrophysics and Supercomputing, Swinburne University of
Technology, Hawthorn, Victoria 3122, Australia\\}

\begin{abstract}
We propose that synchronized triggering of star formation in
gas-rich galaxies is possible during major mergers
of  cluster of  galaxies,
based on new numerical simulations of the time evolution
of the physical properties of the intracluster medium (ICM)
during such a merger event. Our numerical simulations show
that the external pressure of the ICM in which cluster member
galaxies are embedded, can increase significantly during cluster merging.
As such, efficient star formation can be triggered in gas-rich
members as a result of the strong compression of their cold gas
by the increased  pressure.
We also suggest that these star-forming galaxies can subsequently be transformed into
poststarburst  galaxies, with their spatial distribution within the cluster
being different to the rest of its population.
We discuss whether this possible merger-induced enhancement in the
number of star-forming and post-star-forming cluster galaxies is
consistent with the observed evolution of galaxies
in merging clusters.
\end{abstract}

\keywords{
galaxies:halos --
galaxies:structure --
galaxies:kinematics and dynamics --
galaxies:  evolution 
}

\section{Introduction}

Recent $X$-ray observations have revealed that some clusters of
galaxies have ``cold fronts'' which may have been formed
as a result of them undergoing a major/minor  merger (Owers et al. 2009a, b).
These observations, as well as those of substructures
in clusters (e.g., Forman \& Jones 1990; Briel et al. 1991;
Escalera et al. 1994), 
strongly suggest that a significant fraction of clusters
might have experienced merger events. There are also observations
which suggest that cluster merging affects the global star formation
in cluster member galaxies (e.g., Caldwell et al. 1993;
Caldwell \& Rose 1997;  Miller et al. 2003; Ferrari et al. 2005).
These observations beg the key question 
as to how and why some merging clusters show larger fractions of
starburst and  post-starburst galaxies than others 
(e.g., Caldwell \& Rose 1997, Owen et al. 2005).

From a theoretical viewpoint, it is unclear whether and how cluster merging
can significantly change the number fractions of starburst and poststarburst   
galaxies. Bekki (1999) showed that the time-dependent tidal fields of merging
groups and clusters of galaxies can trigger secondary starbursts in their member
galaxies and thus change the number fractions of these galaxies.
Fujita et al. (1999) showed that the star formation rates of galaxies
during a major cluster merger can decrease because of ram-pressure stripping
of the interstellar gas initially within the galaxies.
 As such, the fractions of
blue, actively star-forming galaxies can decrease and then 
the fractions of poststarburst galaxies can increase.
Recent numerical simulations have shown
that  strong ram  pressure from the ICM can significantly increase
the star formation rates in cluster galaxies (Bekki \& Couch 2003; Kronberger et al.
2008). Therefore, it is timely to revisit the question
as to whether the rapidly evolving state of the ICM within a merging cluster can
significantly change the number fractions of starburst and poststarburst galaxies.

The purpose of this Letter  is to thus show, for the first time,
that cluster merging has the potential to trigger  
star formation among a significant fraction of the member galaxies
{\it in a simultaneous way}:
this ``synchronized'' activity might be an important clue to better 
understanding and discriminating merger-driven galaxy evolution in 
cluster galaxy populations. 
We investigate the orbital evolution of cluster member galaxies
and the external pressure of the ICM surrounding the galaxies
during the cluster merging phase, in order to determine 
how such a dynamical event might influence the star formation histories of 
cluster populations. Our previous simulations showed that if the pressure of
the ICM becomes sufficiently high, it can trigger the 
collapse of giant molecular clouds (GMCs) and hence bursts of
star formation in galaxies (Bekki \& Couch 2003; see also
Kronberger et al. 2008  for
ram-pressure-induced star formation).
We thus adopt a model in which the star formation rates of galaxies
in merging clusters are significantly increased if the pressure ($P$) 
of the ICM surrounding the galaxies exceeds the internal pressure of 
the GMCs.

\section{The model}

In order to simulate the time evolution of dark matter halos
and the  ICM in merging clusters, 
we use the latest version of GRAPE
(GRavity PipE, GRAPE-7), which is the special-purpose
computer for gravitational dynamics (Sugimoto et al. 1990).
We use our original GRAPE-SPH code (Bekki \& Chiba 2006; Bekki 2009)
which combines
the method of smoothed particle
hydrodynamics (SPH) with GRAPE for calculations of three-dimensional
self-gravitating fluids in astrophysics. 
Since the models for the structures of dark matter halos 
and the physical properties of hot gas within the halos
are already given in detail by Bekki (2009),
we only briefly describe the models here.

The structure of each of the two clusters in a merger is modeled 
using an ``NFW'' profile predicted by
the cold dark matter cosmology (Navarro et al. 1996),
and the masses and sizes of the clusters are represented by
$M_{\rm cl}$ and $R_{\rm cl}$, respectively.
Henceforth, all masses and lengths are measured in units
of $M_{\rm cl}$ and $R_{\rm cl}$, respectively, unless otherwise
specified. Velocity and time are measured in units of $V_{\rm cl}$ = $
(GM_{\rm cl}/R_{\rm cl})^{1/2}$ and $t_{\rm cl}$ = $(R_{\rm
cl}^{3}/GM_{\rm cl})^{1/2}$, respectively, where $G$ is the
gravitational constant and assumed to be 1.0 in the present study.
These $V_{\rm cl}$ and $t_{\rm cl}$ correspond to
the circular velocity and dynamical time scale at $R_{\rm d}$, respectively.

The $c$ parameter ($=r_{\rm s}/r_{\rm vir}$, 
where $r_{\rm s}$ and $r_{\rm vir}$ 
are the scale and virial radii of the NFW profile, respectively)  for a cluster
with $M_{\rm cl}$ (=$M_{\rm dm}$)
is chosen according to the predicted $c$-$M_{\rm dm}$ relation
in the $\Lambda$CDM simulations (e.g., Neto et al. 2007).
A reasonable value of $c$ is thus 4.7 for 
$M_{\rm dm}=10^{14} {\rm M}_{\odot}$.
The larger and smaller clusters in a merger, whose
mass ratio is denoted as $m_{\rm 2}$ ($0.1 \le m_{\rm 2} \le 1$), 
are referred to as
CL1 and CL2, respectively, for convenience.
If CL1 has mass $M_{\rm cl}$ and radius $R_{\rm cl}$, then
CL2 has mass $m_{2} M_{\rm cl}$ and radius $\sqrt{m_2}R_{\rm cl}$.

The ICM  has mass $M_{\rm g}$
and the same spatial distribution (${\rho}_{\rm g}$) as the dark
matter and is assumed to be initially in hydrostatic equilibrium. The
initial gaseous temperature of an ICM particle is therefore determined by
the gas density, total mass, and gravitational potential at the location of the
particle via Euler's equation for hydrostatic equilibrium (e.g., equation
1E-8 in Binney \& Tremaine 1987). 
Therefore  gaseous temperature $T_{g}(r)$ at radius $r$ from the center
of a cluster  can be described as:

\begin{equation}
T_{\rm g}(r)=
\frac{m_{\rm p}}{k_{\rm B}}
\frac{1}{{\rho}_{\rm g}}
\int_{r}^{\infty}
{\rho}_{\rm g}(r)
\frac{GM(r)}{r^2}
dr,
\end{equation}
where $m_{\rm p}$, $G$, and $k_{\rm B}$ are the proton mass,
the gravitational constant, and the Boltzmann constant, respectively,
and $M(r)$ is the total mass within $r$ determined by
the adopted mass distributions of dark matter and baryonic
components in the cluster.
Radiative cooling is not included in the present study so that
the hydrodynamical equilibrium of halo gas can be obtained
from isolated cluster  models.
We can expect  that the shocked gaseous regions 
in models with  radiative cooling
have  significantly higher gas densities and pressure,
and therefore galaxies in the models can be more strongly
influenced by cluster merging during their passage of the shocked regions.
We adopt a representative  value of $M_{\rm g}=0.136 M_{\rm cl}$
(e.g., McCarthy et al. 2008).

Galaxies in a cluster are represented by collisionless particles
and their spatial distribution follows the NFW profile
with $c=3$.  The canonical Schechter function
is adopted for generating a galaxy luminosity/mass function for
luminosities ranging from $0.01L^{\ast}$ to $2.5L^{\ast}$
in a cluster.
The total mass (thus number)  of galaxies in a cluster
with $M_{\rm cl}$ is determined by the mass-to-light-ratio, 
that itself is dependent on $M_{\rm cl}$
(Marinoni \& Hudson  2002). Therefore, as an example,
a cluster with $M_{\rm cl}=10^{14} {\rm M}_{\odot}$
has 114 galaxies. 
Cluster member  galaxies in a cluster have an isotropic
velocity dispersion just as the  dark matter of the cluster does. 

Galaxies in the present study
have no halo gas that might well shield ISM of disk galaxies
and thus weaken the possible physical effects of ICM (e.g., triggering
star formation) on the ISM. Previous numerical simulations, however,
showed that halo gas around disk galaxies can be efficiently stripped
by ICM even in isolated clusters (e.g., Bekki 2009): the halo gas
would not significantly weaken the effects of ICM on galaxy evolution
in merging groups and clusters.
The ISM of galaxies is not included in the present models so that
the details of star formation caused by cluster merging can not be 
properly investigated. The present study thus assumes that
if pressure of ICM around galaxies can become high enough, 
then the star formation histories
can be significantly changed irrespective  of galaxy properties. 
This assumption would be reasonable, if  GMC properties do not depend
strongly on  galaxy ones.

The relative position and velocity of CL2 with respect to CL1 are
described as ($X_{\rm r}$, $Y_{\rm r}$, $Z_{\rm r}$)
and ($U_{\rm r}$, $V_{\rm r}$, $W_{\rm r}$), respectively.
For all models described in the present study,
$X_{\rm r}=2R_{\rm cl}$ 
(where $R_{\rm cl}$  is the size of CL1), $Z_{\rm r}=0$, $U_{\rm r}=0$,
and  $W_{\rm r}=0$. Therefore $Y_{\rm r}$ is the impact parameter
of cluster merging and $V_{\rm r}$ is the initial relative velocity
of merging two clusters.
Cluster merging processes and subsequent ICM evolution depend strongly
on the four parameters, $M_{\rm cl}$, $m_{2}$,  $Y_{\rm r}$,
and  $V_{\rm r}$. Although we have conducted a large parameter study,
we only show the results of the models with 
$M_{\rm cl}=10^{14} {\rm M}_{\odot}$,
$0.1\leq m_{2} \leq 1$, 
$Y_{\rm r}=0.5 R_{\rm cl}$,
and $V_{\rm r} = -V_{\rm cl}$.
Parameter dependencies of the results will be described 
in detail in our forthcoming papers
(Bekki et al. 2009).

Here we focus in particular on the ``standard model'' with
$M_{\rm cl}=10^{14} {\rm M}_{\odot}$,
$m_{2}=0.25$,
$R_{\rm cl}=2.0$ Mpc,
$V_{\rm cl}=595$ km s$^{-1}$,
$t_{\rm cl}=2.0$ Gyr,
$Y_{\rm r}=1.0$ Mpc,
and $V_{\rm r} = -595$ km s$^{-1}$,
because it shows one of the typical behaviors of the time evolution 
of the external gas pressure around  galaxies in merging clusters.
The total particle number of dark matter and gas particles
used in a cluster merger depends on $m_{2}$:
it is  264,000 for $m_2=0.25$ and $440,000$ for $m_2=1$.
We also run an ``isolated model'' in which a cluster evolves
without merging with any other clusters. Throughout this paper, 
the time $T$ represents the elapsed time since the start of the 
simulation.

We mainly investigate the time evolution of the external pressure
($P$) of the ICM surrounding galaxies,
as first done by Evrard (1991).
We investigate $P$ (static pressure) at each time step
for each galaxy during  $4t_{\rm cl}$ evolution of cluster merging
and thereby estimate the maximum value ($P_{\rm max}$) for each galaxy
and the time ($t_{\rm max}$)  when $P=P_{\rm max}$.
We then check whether $P_{\rm max}$ is larger than
the threshold pressure ($P_{\rm thres}$) required to induce
the global collapse of giant molecular clouds to form massive star clusters
(e.g., Elmegreen \& Efremov 1997).
We set $P_{\rm max}$ to be  $2 \times 10^5$ $k_{\rm B}$ where  $k_{\rm 
B}$ is Boltzmann's constant  (e.g., Bekki et al. 2002).

\section{Result}

Figure 1 shows how the distributions of galaxies for the larger
(CL1) and smaller clusters (CL2)
can change during cluster merging with $m_{2}=0.25$.
During the strong hydrodynamical interaction between the ICM of 
the merging clusters ($T \approx 3$ Gyr),  
some fraction of the galaxies, in particular, those within CL2,
pass through the high-density, high-pressure region where the 
two gas spheres collide.
The CL2 group persists as a coherent substructure 
in the spatial distribution of galaxies 
during the final dynamical relaxation phase of cluster
merging ($T \approx 5$ Gyr).
The galaxies from CL1 and CL2 are finally  well mixed 
and show no clear differences in their spatial distributions
with respect to the center of the newly formed cluster
($T=7$ Gyr).

Figure 2 compares the time evolution of $P$
(pressure of ICM surrounding  galaxies)
in the standard merger model with that in the isolated model 
for the same selected galaxy. 
Figure 2 clearly shows that $P$ dramatically increases
owing to the passage of the galaxy through the high-pressure
shocked region of the merging cluster
when the clusters collide ($T\approx 3$ Gyr). $P$ does not change
much at all in the isolated model.
The simulated $P$ exceeds the threshold pressure $P_{\rm th}$
($=2.0\times 10^5 k_{\rm B}$) for the collapse of GMCs, 
so that efficient star formation within GMCs is highly likely.
This clearly demonstrates that cluster merging can strongly
influence the star formation activity within their member galaxies.

Figure 3 shows that the distribution of $t_{\rm max}$ for CL2  has a
peak around $T \approx 3$ Gyr, whereas CL1
has a peak at $T \approx 4$ Gyr. 
For CL2, the number fraction
of galaxies with $t_{\rm max}$  ($F_{\rm max}$)
for  $T \approx 3$ Gyr 
can be as high as 0.4, which implies that 
a significant fraction of galaxies in CL2 can experience
strong pressure  by the surrounding ICM
{\it simultaneously}
when two clusters collide.
Figure 3 also shows that 
$P_{\rm max}$ 
for most  of galaxies  with $t_{\rm max} \approx 3$ Gyr
can be higher than $P_{\rm thres}$: some of them also
show  $P_{\rm max}$ larger than 
$10^6 k_{\rm B}$.
These results suggest that
a significant fraction of galaxies in merging clusters
can be simultaneously influenced by the dramatically  increased
external  pressure
of ICM.

\section{Discussion and conclusions}

\subsection{Possible distributions of starburst and poststarburst 
galaxies}

In order to discuss possible distributions
of starburst and poststarburst galaxies at $T$ in each simulation,
we assume  that galaxy particles
with  $P_{\rm max} \ge P_{\rm thres}$  that have
$T-t_{\rm max} \le 0.1$ Gyr and 0.1Gyr $\le T-t_{\rm max} \le$ 1Gyr
are labeled as starburst galaxies (SBCs) and poststarburst
galaxies (PSBCs),
respectively.
Given that previous numerical simulations confirmed the formation
of ``E+A''  galaxies (``E+A''s)  from starburst galaxies  less than 1 Gyr after
strongly secondary starbursts (e.g., Bekki et al. 2005),
the above criterion can be reasonable.

Figure 4 shows 
the distribution of PSBCs  
at $T=4$ Gyr appears to have a weakly  elongated structure (or substructure)
in the direction of $X=-1$ Mpc and $Y=-0.5$ kpc.
This reflects the fact that CL2 has not been dynamically  relaxed
completely yet 
and some PSBCs are still within CL2.
The number fraction of PSBCs ($f_{\rm PSBC}$)
becomes large ($\sim 0.22$) at $T=4$ Gyr,
mainly because a large fraction of galaxies pass through
the high pressure region of the merging cluster around $T=3$ Gyr.
The main reason for the
apparent lack of PSBCs in the core of  CL1 (i.e., $|X| \le 0.2$ Mpc
and ($|Y| \le 0.2$ Mpc) is that galaxies experienced
strong external pressure from the ICM there some 0.9-1.0 Gyr ago
(i.e., at $T=3.0 - 3.1$ Gyr). 
The distribution of PSBCs changes significantly as the merging clusters
dynamically relax.

The fraction of galaxies that experience
high external pressure with $P_{\rm max} \ge P_{\rm thres}$
depends on  $m_{2}$, 
such that it is likely to be higher in models with larger $m_{2}$:
it is 0.29 for $m_{2}=0.1$ and 0.74 for  $m_{2}=1.0$.
The fraction of galaxies that experience
very high external pressure with $P_{\rm max} \ge 5P_{\rm thres}$
depends strongly on  $m_{2}$: it is only 0.008 for $m_{2}=0.1$
and 0.51 for $m_{2}=1.0$. The main reason for these
dependencies is that major merging can form
more strongly  shocked gaseous regions over a larger volume
of the merging clusters, so that a larger number of
their member galaxies pass through the shocked regions
during the merger.

Just after cluster merging, the PSBCs appear to be 
distributed in a ring-like structure, particularly in the 
case of the major merger 
models with $m_{2}=1$. This is due to the strong central concentration 
of SBCs in the merger remnants: all galaxies in the core are identified
as SBCs, which is just due to the model assumption: galaxies are always
identified
as SBCs irrespective of gas content
(i.e., whether they are gas-poor, poststarburst galaxies)
whenever they are in the core region where the external pressure
is high enough to trigger SBs. 
Thus, it is not 
appropriate to use the results of this paper
to discuss the observational results of Owers et al. (2010), which show an 
intriguing distribution of PSBCs in clusters that have 
undergone a recent major merger,

\subsection{Larger fractions of starburst galaxies in merging clusters ?}

We have shown that cluster merging can dramatically increase 
the pressure of the ICM surrounding the cluster member galaxies, 
to the extent that
a significant fraction of
the galaxies can be simultaneously affected. 
Previous numerical simulations have shown that 
GMCs in gas-rich galaxies that are exposed to this increased pressure 
of the ICM are strongly  compressed, thereby triggering efficient 
star formation within them in their high-density regions (Bekki \& Couch 2003).
We therefore suggest that (i)\,cluster merging can trigger starbursts
in gas-rich galaxies embedded within those regions of the ICM 
that have undergone this dramatically increase in pressure, 
and (ii)\,these starbursts can occur simultaneously for a significant 
fraction of the galaxies within the 
merging clusters.

It should be stressed, however, that ram pressure stripping can become 
much more effective during cluster merging so that cold HI gas 
within disk galaxies and their halos -- which can fuel star formation 
-- can be efficiently stripped
(Fujita et al. 1999; Bekki 2009). This ram pressure stripping of HI gas
would cause severe truncation of star formation after the GMCs 
are converted into new stars during cluster merging. This process 
would lead naturally to the formation of PSBC's (also known as 
``E+A'' galaxies). Therefore, cluster merging can provide a 
possible explanation for the observed larger fraction of
E+A galaxies in some clusters with substructures (e.g., Caldwell \& Rose 1997).

Whether major starbursts are triggered in disk galaxies as a 
result of the strong external pressure exerted by the ICM 
during cluster merging, depends strongly on the gas mass 
fractions ($f_{\rm g}$) within their disks.
It is well known that higher redshift clusters of galaxies have 
a larger fraction of blue galaxies than their lower redshift counterparts
(e.g., Butcher \& Oemler 1978). This likely means that there 
is a larger fraction of gas-rich galaxies in higher redshift clusters, 
although only future HI observations can verify this directly. 
The present result thus implies that synchronized formation of 
starburst galaxies is more likely to occur in higher redshift merging 
clusters of galaxies.

Galaxies located in the core regions of clusters can 
experience strong ram pressure stripping of their ISM 
by the ICM, to the extent that they lose most of 
their ISM gas (i.e. $f_{g}\sim 0$). They are therefore unlikely 
to experience starbursts during cluster merging, because they 
are already gas-poor prior to cluster merging. Thus the present 
study is likely to overestimate the fractions of starburst 
galaxies to some extent. We need to model properly the 
variation in $f_{\rm g}$ in galaxies of different Hubble type, 
and how that translates into a varying $f_{\rm g}$ with radius 
from the cluster centre, in order to predict much more precisely 
the possible fractions of starburst and poststarburst 
galaxies in merging clusters in our future studies.

It should also be noted that it is not only the external pressure 
of the ICM in clusters that can trigger starbursts in gas-rich 
galaxies (e.g., Kronberger et  al. 2008), but also time-dependent 
cluster tidal fields (Bekki 1999). 
It therefore appears inevitable that merging clusters are likely 
to have a larger fraction of starburst galaxies than non-merging clusters. 
It would be difficult, however, for observational studies to 
determine whether tidal effects or increased external ICM 
pressure is the main driver for such an increased starburst 
fraction. Since the time-varying tidal fields in merging 
clusters can also transform {\it stellar} disks (Bekki 1999), 
whereas the external pressure
of the ICM is unlikely to do so, the morphological properties 
of each individual
starburst galaxy will provide important clues as to which 
of the above two effects are responsible for their origin.

Recent and ongoing photometric and spectroscopic observations 
of galaxy properties in a large sample of clusters with and 
without cold fronts will soon reveal the number fractions of 
starburst and poststarburst galaxies and their spatial 
distributions and kinematics in these possibly merging and 
non-merging clusters (e.g., Hwang \& Lee 2009; Ma et al. 2010;
Owers et al. 2010). These statistical 
studies will enable us to compare the simulated kinematics 
(e.g., line-of-sight velocity distributions) of
starburst galaxies with those observed, thereby allowing 
more robust conclusions to be drawn as to whether the 
star formation histories of cluster member galaxies are 
dramatically changed by cluster merging. The observed peculiar 
spatial distributions (e.g., ring-like structures) of poststarburst 
galaxies in clusters with substructures (e.g., the Coma cluster; 
Poggianti et al. 1999) will place strong constraints on the 
mass-ratios and radial velocities of merging clusters.

\acknowledgments
KB, MSO and WJC all acknowledge the financial support of the 
Australian Research
Council throughout the course of this work. Numerical computations 
reported here were carried out both on the GRAPE system at the 
University of Western Australia  and on those kindly made available 
by the Center for computational astrophysics
(CfCA) of the National Astronomical Observatory of Japan.

\newpage

\begin{figure}
\epsscale{.60}
\plotone{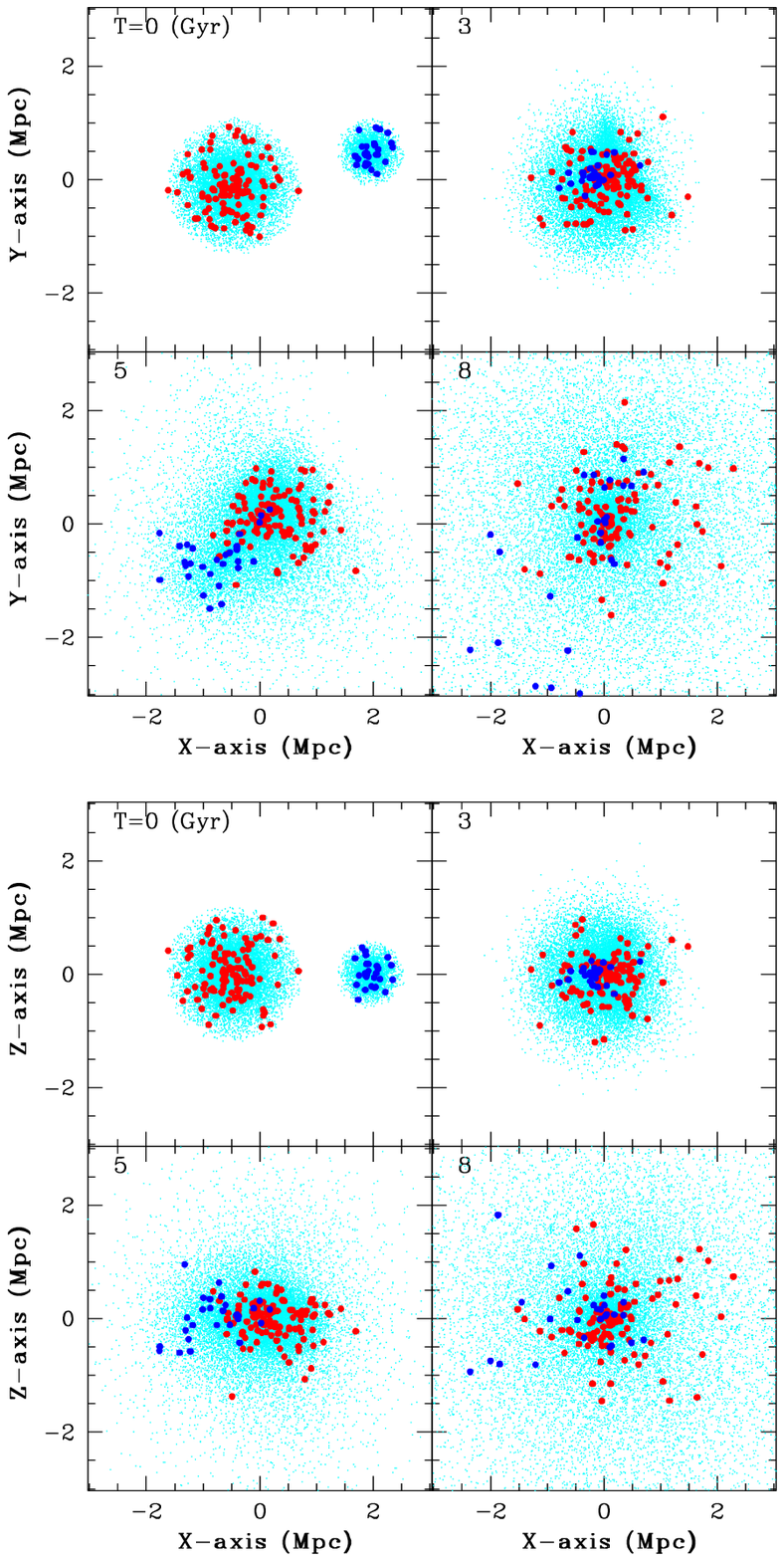}
\figcaption{
Time evolution of the spatial distributions of the ICM (smaller cyan dots),
the galaxies in CL1 (bigger red dots), and the galaxies
in CL2 (bigger blue dots), projected onto the $x$-$y$ (upper four panels) and
$x$-$z$ planes (lower four panels), for the standard model.
$T$ shown in the upper left corner of each panel
represents the time that has elapsed since the start of the simulations.
\label{fig-1}}
\end{figure}

\begin{figure}
\epsscale{0.8}
\plotone{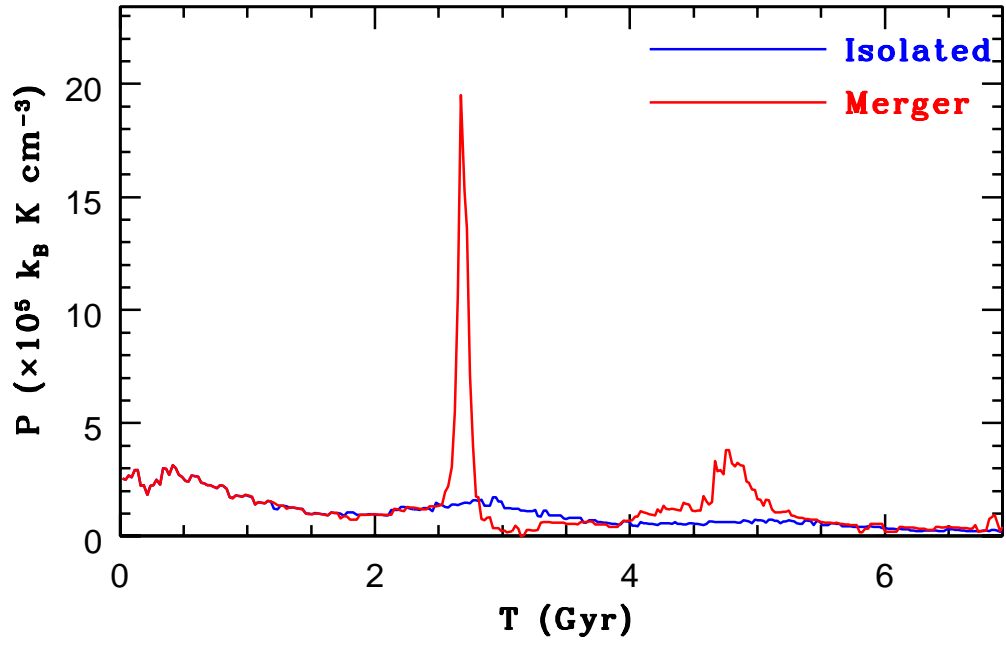}
\figcaption{
Time evolution of $P$ for a selected galaxy particle for the isolated
model (blue),  and the standard merger model (red).
\label{fig-2}}
\end{figure}

\begin{figure}
\epsscale{0.7}
\plotone{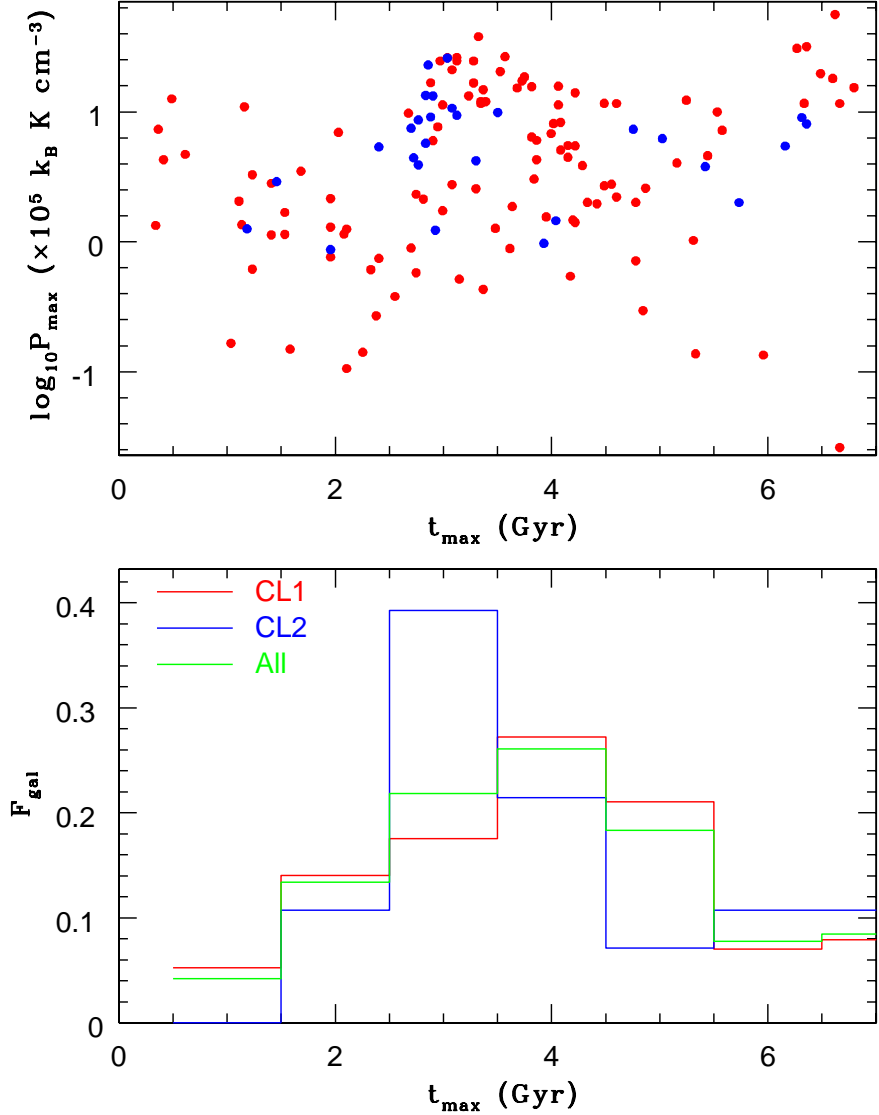}
\figcaption{
The upper panel shows
the distribution of galaxies in CL1 (red) and CL2 (blue)
on the log\,$P_{\rm max}-t_{\rm max}$ plane (upper) for the standard model.
The lower panel shows
the $t_{\rm max}$ distribution for CL1 (red), CL2 (blue)
and all galaxies in CL1 and CL2 (green), expressed in terms of
the  number fraction ($F_{\rm gal}$) of galaxies at each  $t_{\rm max}$ bin.
The presence of galaxies showing large $P_{\rm max}$
($>10^6 {\rm k}_{\rm B}$ K cm$^{-3}$) at $t_{\rm max}<1$ Gyr
is due largely
to the fact that they can pass through the very inner  region of the cluster
(at their early orbital evolution phases)
where the static pressure is rather high.
\label{fig-3}}
\end{figure}

\begin{figure}
\epsscale{0.7}
\plotone{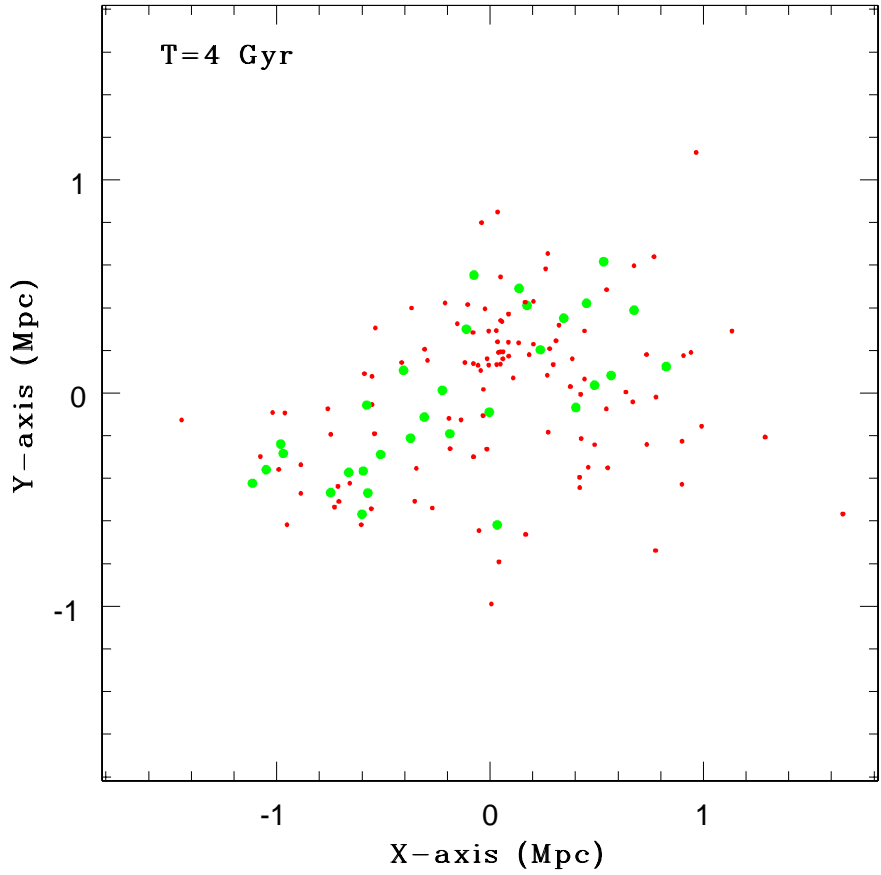}
\figcaption{
The spatial distributions of PSBCs (bigger green dots) and non-PSBCs (smaller
red dots) projected onto the $x$-$y$ plane
at $T=4$ Gyr ($\sim 1$ Gyr after the two cluster violently collide)
for the standard model.
\label{fig-4}}
\end{figure}

\end{document}